\documentclass[aip,jcp,preprint]{revtex4-1}
\usepackage{amsmath}
\usepackage{amsfonts}
\usepackage{bm}
\usepackage{braket}
\usepackage{mathrsfs}
\usepackage{dcolumn}
\usepackage{graphicx}
\usepackage{epstopdf}
\usepackage[colorlinks=true,linkcolor=black,anchorcolor=black,citecolor=black,filecolor=black,menucolor=black,runcolor=black,urlcolor=blue]{hyperref}
\usepackage{dcolumn}
\newcommand{\JPA}{{\it J. Phys. A: Math. Gen.} } 
\newcommand{\JMP}{{\it J. Math. Phys.} } 
\newcommand{\jcp}{J.~Chem.~Phys.}

\DeclareMathOperator{\res}{Res}
\newcommand\Ord{\Or}
\newcommand\Us{\hat{\boldsymbol{\mathsf  U}}\vphantom{\boldsymbol{\mathsf  U}}}

\newcommand\Uo{\hat{ U}\vphantom{ U}}
\newcommand\Vo{\Uo^\dagger}
\newcommand\Vs{\hat{\bm{\mathsf  V}}}
\newcommand\expp[1]{{\rm exp}_{+}(\ \rmi\int_0^\alpha{#1}\,\rmd s)}
\newcommand\expm[1]{{\rm exp}_{-}(-\rmi\int_0^\alpha{#1}\,\rmd s)}
\newcommand\vecp[1]{\vec{#1}}

\newcommand\Ls{{\hat{\bm{\mathcal L}}}}
\newcommand\Ps{\hat{\bm{\mathcal  P}}}

\newcommand\Ss{\hat{\bm{\mathcal  S}}}
\newcommand\Rs{\hat{\pmb{\mathscr  R}}}
\newcommand\Srs{\hat{\bm{\mathcal  R}}}
\newcommand\Ds{\hat{\bm{\mathcal  D}}}
\newcommand\Zs{\hat{\bm{\mathcal  Z}}}

\newcommand\ac[2]{\hat a_{#1}^{\dagger {#2}}}
\newcommand\xac[2]{\hat a_{#2}^{\dagger}\vphantom{\hat a}^{#1}}
\newcommand\xa[2]{\hat a_{#2}^{{#1}}}
\newcommand\ih{\frac{\rmi}{\hbar}}
\newcommand\hi{\frac{\hbar}{\rmi}}
\newcommand\Ssh{\Ss}
\newcommand\Psh{\Ps}
\newcommand\Lsi{\Ls_{\Hi}}
\newcommand\Hi{\hat H_1}
\newcommand\rmi{i}
\newcommand\rme{e}
\newcommand\rmd{d}
\newcommand{\Or}{\mathord{\mathrm{O}}} 
\newcommand{\eref}[1]{(\ref{#1})} 
\begin{document}
\title[Kato  expansion in quantum canonical perturbation theory]{Kato  expansion in quantum canonical perturbation theory
}

\author{Nikolaev~A.\,S.}
\affiliation{  Institute of Computing for Physics and Technology, Protvino, Moscow reg., Russia}
\homepage{{http://andreysnikolaev.wordpress.com}}
\affiliation{  RDTeX LTD, Moscow, Russia}
\email{Andrey.Nikolaev@rdtex.ru}

\begin{abstract}
This work  establishes a connection between canonical perturbation series in quantum mechanics and a Kato  expansion for the resolvent of the Liouville superoperator. 
Our approach leads to  an  explicit expression  for  a generator of a block-diagonalizing Dyson's ordered exponential 
 in arbitrary perturbation order.
Unitary intertwining of perturbed and unperturbed averaging superprojectors  allows for a description of  
  ambiguities in the generator and block-diagonalized  Hamiltonian.
  
  We   compare the  efficiency of the corresponding computational algorithm with the efficiencies
  of the Van Vleck and Magnus methods for high perturbative orders.
\end{abstract}

\maketitle

\section{Introduction.}

Canonical Perturbation Theory (CPT) 
  was historically the first perturbation theory in quantum mechanics \cite{3man}.
It    iteratively constructs  a unitary transformation 
of the Hamiltonian matrix into a block-diagonal form.
 The resulting {\em effective}   Hamiltonian \cite{soliverez81,jorgensen75} 
then  allows for a  solution of the eigenvalue problem.

Nowadays,  Van Vleck  \cite{VanVleck} CPT  in Primas'  superoperatorial formalism \cite{Primas} is  widely used   in quantum chemistry and molecular spectroscopy \cite{Klein,Tyuterev}.
Contemporary precise  models demand time-consuming computations up to very high perturbative orders \cite{tyuterevhigh}.
This is why new efficient  algorithms are  important.

We present here a noniterative construction of the  block-diagonalizing   transformation for CPT. 
Our approach follows the recent application of Kato perturbation expansion to classical mechanics \cite{nikolaev4}.
It uses the remarkable analogies between mathematical formalisms of perturbation expansions in classical and quantum mechanics.

 The method is  based on a Kato series \cite{Kato} for the 
 the Laurent coefficients of the 
  resolvent of the Liouville superoperator and the Dyson's ordered exponentials.
We demonstrate a  regular structure in  perturbation series and 
derive  an  explicit expression  for the  generator of a block-diagonalizing transformation in any perturbation order:
\[\hat G = \rmi \,\Ss_{\hat H} \Hi.\]
Here, $\Ss_H$  is the  partial pseudo-inverse of the  perturbed Liouville superoperator.
    Unitary intertwining of perturbed and unperturbed averaging superprojectors  allows for a description of  
  ambiguities in the generator and block-diagonalized  Hamiltonian.
The corresponding  perturbative algorithms are sufficiently efficient even for high-order computations.

Note that all calculations here are only formal, in a sense that neither a discussion of power
series convergence nor conditions for the existence of constructed superoperators are present.

The downloadable 
``Supplementary data files'' \cite{Demo} contain demonstrations  and large formulae.
The demonstrations use the freeware computer algebra system FORM \cite{Form}.

\section{Basic perturbation superoperators.} \label{sec:basic}
Consider the  Hamiltonian $\hat H$
that differs from an exactly solvable one by a  perturbation:
\[
\hat H=\hat H_0+\alpha \Hi.
\]
We will discuss here only time-independent Hamiltonians that have a discrete spectrum.

The goal of canonical perturbation theory is to transform  the perturbed Hamiltonian into a simpler block-diagonal operator 
by a near-identity unitary   transformation. 
Such transformations may be  handled conveniently using  superoperator formalism \cite{Primas,Lowdin}. 
Here we will  outline only some basics of it. 

For any  $\hat G$, one may introduce  a quantum Liouville superoperator  acting on the operator space:
\[
\Ls_{\hat G} = [\hat G,\,.\,], \qquad \Ls_{\hat G} \hat F = [\hat G,\hat F]. 
\]
We will use the bold calligraphic-style  letters here to denote the superoperators.
Due to the  Jacobi identity,  the Liouvillian is the ``derivation'' of a commutator: 
\[
\Ls_{\hat G} [\hat F, \hat H]= [\Ls_{\hat G} \hat F,\hat H ]+[ \hat F,\Ls_{\hat G} \hat H]. 
\]

The $\Ls_{\hat H_0}$ of the discrete spectrum system
admits the decomposition of its domain  into the
direct sum $\mathfrak{ N} \oplus \mathfrak{ R}$, where $\mathfrak{ N}$ or $\mathfrak{R}$
 are the kernel and the range space, respectively. 
In order to realize this decomposition, Primas \cite{Primas} introduced the following \emph{averaging} ({block-diagonalizing}) superoperator:
\begin{equation}
\Ps_{\hat H_0} \hat F =  \lim\limits_{\lambda\to +0}\lambda\int_0^{+\infty}\rme^{-\lambda t} \rme^{\ih t \hat H_0}\,\hat F\, \rme^{-\ih t \hat H_0}  \rmd t 
=  \lim\limits_{\lambda\to +0}\lambda\int_0^{+\infty} \rmd t \,\rme^{-\lambda t} \rme^{\ih t\, \Ls_{\hat H_0}} \hat F.  \label{Pdef}
\end{equation}
This  superoperator is the projector $\Ps_{\hat H_0}^2 = \Ps_{\hat H_0}$. It projects the operator $\hat F$ onto  the kernel space 
$\mathfrak{N}$ of block-diagonal operators commuting with $\hat H_0$. 

The complementary projector $1-\Ps_{\hat H_0}$ extracts the non-commuting with the $\hat H_0$  (off-diagonal) part of the $\hat F$. 
It projects on the range space $\mathfrak{R}$  where the inverse of $\Ls_{\hat H_0}$ exists.
This inverse is the \emph{integrating superoperator} $\Ss_{H_0}$,  which is also called  the solution of the homological equation, tilde 
operation, zero-mean antiderivative,  
Friedrichs ${\bf \widehat\Gamma}$ operation, ${\bf \frac{1}{k}}$ operator, division 
operation, etc. Its invariant definition by Primas  \cite{Primas} is as follows: 
\begin{equation}
\Ss_{\hat H_0} = -\ih\  \lim_{\lambda\to{+0}}\int_0^\infty \rmd t \,\rme^{-\lambda t} \rme^{\ih t\,\Ls_{\hat H_0}} \left(1-\Ps_{\hat H_0}\right).\label{Sdef}
\end{equation}
It is easy to check that:
\begin{align*}
\Ls_{\hat H_0}\Ss_{\hat H_0} &= \Ss_{\hat H_0}\Ls_{\hat H_0}=1-\Ps_{\hat H_0},\\
\Ss_{\hat H_0}\Ps_{\hat H_0} &= \Ps_{\hat H_0}\Ss_{\hat H_0}\equiv 0.
\end{align*}

The superoperators $(\Ls_{\Hi}$, $\Ps_{\hat H_0}$,  $\Ss_{\hat H_0})$ are the building blocks of the canonical perturbation series \cite{Primas}.
Quantum mechanics uses the following methods for  their computation: 
\begin{itemize}
\item The {\em  Energy Representation} of an unperturbed Hamiltonian provides the eigenbasis of 
the Liouville superoperator \cite{Lowdin}.
The action of the basic superoperators 
on any operator $\hat F$
 can be directly computed in this representation  as follows:
\[
\Ps_{\hat H_0} \hat F =\sum_{E_m = E_n} F_{mn} \ket{m} \bra{n},\qquad
\Ss_{\hat H_0} \hat F =\sum_{E_m \ne E_n} \frac{F_{mn}}{E_m - E_n} \ket{m} \bra{n}.
\]
These  well-known expressions are used frequently as the definitions of $\Ps_{\hat H_0}$ and $\Ss_{\hat H_0}$.

\item The method of {\em   Contact Transformations} \cite{PhysRev.56.895}  (quantum Birkhoff-Gustavson normal form  \cite{Birkhoff,Gustavson}) applies to the perturbed harmonic oscillator
\[\hat H = \sum_{k=1}^{\textbf d} \frac{\omega_k}{2} \left({\hat p}_k^2 +   {\hat q}_k^2\right) + \alpha \Hi({\hat p,\hat q}).\]
It uses the  representation of  ladder  operators:
\[\hat q_k  = \sqrt{\frac{\hbar}{2}} (\hat a^\dagger_k + \hat a_k),\qquad p_k = \rmi\sqrt{\frac{\hbar}{2}}  (\hat a^\dagger_k - \hat a_k),\qquad [\hat a_k, \hat a_l^\dagger]=\delta_{kl}.
\]
The action of basic superoperators  on a Wick-ordered polynomial operator
$\hat F({\hat  p,\hat  q})= \sum \tilde F(\vecp m,\vecp n) \, \ac{}{\vecp m}\hat a^{\vecp n}$ is  as follows:
\[
\Ps_{\hat  H_0} \hat F = \sum_{(\vecp\omega,\vecp m - \vecp n) =0} \tilde F(\vecp m,\vecp n)\, \ac{}{\vecp m}\hat a^{\vecp n},
\qquad
\Ss_{\hat  H_0} \hat F = \sum_{(\vecp\omega,\vecp m - \vecp n)\ne 0} 
\frac{\tilde F(\vecp m,\vecp n)}{\hbar(\vecp\omega,\vecp m - \vecp n)}\, \ac{}{\vecp m}\hat a^{\vecp n}.
\]

\end{itemize}

\section{The superresolvent}

Consider the resolvent of the Liouville superoperator:
\begin{equation}
\Rs_{H}(z) =\frac{1}{{\ih \Ls_{H}-z}}.\label{Res}
\end{equation}
This  superoperator-valued function of the complex variable $z$ is the Laplace transform of  the evolution superoperator  

\[
\Rs_{H}(z) = - \int_0^{+\infty}\rmd t\, \rme^{-z t} \rme^{\ih t\,\Ls_{H}}.
\]

Resolvent singularities are the eigenvalues of the Liouvillian. 
Typically, its  spectrum $\pmb\sigma(\Ls_{\hat H})=\pmb\sigma(\hat H)-\pmb\sigma(\hat H)$  has a richer structure than that of  the corresponding Hamiltonian 
\cite{liu2013}.

 Let us begin with the simple case of   
the pure point spectrum of a quantum system with isolated energy levels.
The singularities of corresponding  Liouvillian superresolvent are located at $E_k-E_m$  and are separable from the origin.

  The existence of 
$\Ps_{\hat H_0}$ 
 and $\Ss_{\hat H_0}$ 
means that the unperturbed superresolvent has a simple pole in $0$. 
The averaging superoperator is its residue in this pole:
\[
\Ps_{\hat H_0} \equiv - \res_{z=0} \Rs_{\hat H_0},
\]
while the integrating superoperator $\Ss_{\hat H_0}$ is its holomorphic part: 
\[
\Ss_{\hat H_0}=\ih \lim_{z\to0}{\Rs_{\hat H_0}(z)(1-\Ps_{H_0})}.
\]
Therefore, the Liouvillian superresolvent combines both basic perturbation superoperators  \cite{nikolaev4}.
This allows us to apply the  powerful  formalism of complex analysis to perturbation theory.  

It is well known \cite{Kato} that, due to the Hilbert resolvent identity 
\[
\Rs_{\hat H}(z_1)-\Rs_{\hat H}(z_2)=(z_1-z_2)\,\Rs_{\hat{H}}(z_1)\,\Rs_{\hat H}(z_2),
\]
the Laurent series for the unperturbed resolvent  has the  form of a geometric progression
\[
\Rs_{\hat H_0}(z)= -{1\over z}\Ps_{\hat H_0} + \sum_{n=0}^\infty z^n  \left(\hi\,\Ss_{\hat H_0}\right)^{n+1}
=\sum_{n=0}^{+\infty} \left(\hi\right)^n \Srs_{H_0}^{(n)}z^{n-1}. 
\]
Here, we have denoted $\Srs_{H_0}^{(0)}=-\Ps_{\hat H_0}$ and $\Srs_{H_0}^{(n)}= \Ss_{\hat H_0}^n$.

The perturbed resolvent  may be more singular. The Laurent series for a general resolvent with an isolated singularity at the origin has
the following form \cite{Kato}:
\[
\Rs_{\hat H} (z)=-\frac{1}{z}\Ps_{\hat H} + \sum_{n=0}^{\infty} z^n \left(\hi\, \Ss_{\hat H}\right)^{n+1}
-\sum_{n=2}^{\infty} z^{-n} \left(\ih\, \Ds_{\hat H}\right)^{n-1}. 
\]
Here, $\Ds_{\hat H}$ is  the eigen-nilpotent superoperator, which does not have an unperturbed analogue.

\subsection{Kato series}
If the  perturbation $\hat H_1$ is relatively bounded with respect to $\hat H_0$, then
the  Rayleigh--Schr\"odinger series converges \cite{Kato} and 
the perturbed superresolvent can be expanded into the  Neumann series as follows:
\begin{align*}
\Rs_{\hat H_0+\alpha \Hi}(z)=&
\Rs_{\hat H_0}-\alpha\ih\, \Rs_{\hat H_0}\Ls_{\Hi}\Rs_{\hat H_0}
+
\alpha^2 \left(\ih\right)^2  \Rs_{\hat H_0}\Ls_{\Hi}\Rs_{\hat H_0}\Ls_{\Hi}\Rs_{\hat H_0}+
\ldots\\
=&\sum_{n=0}^{\infty}  \left(-\alpha\,\ih\right)^n \Rs_{\hat H_0}(z){\left(\Ls_{\Hi}
\Rs_{\hat H_0}(z)\right)}^n.
\end{align*}

The integration  
around a small contour results in the Kato   series \cite{Kato} for the ``perturbed averaging superoperator'':
\begin{align*}
\Ps_{\hat H} &= -{1\over{2\pi \rmi}}\oint_{|z|=\epsilon} \Rs_{\hat H(z)}\rmd z\\
&=-{1\over{2\pi \rmi}}\sum_{n=0}^\infty \oint_{|z|=\epsilon} {\left(-\alpha\ih\right)}^n
\Rs_{\hat H_0}(z){\left(\Ls_{\Hi}\Rs_{\hat H_0}(z)\right)}^n \, \rmd z\\
&={-1\over{2\pi \rmi}}\sum^\infty_{n=0} {\left(-\alpha\ih\right)}^n\!\oint_{|z|=\epsilon}
\left(\sum_{m=0}^\infty {\left(\hi\right)}^m\Srs_{\hat H_0}^{(m)}z^{m-1}\right)
{\left(\Ls_{\Hi}\sum_{k=0}^\infty{\left(\hi\right)}^k\Srs_{\hat H_0}^{(k)}z^{k-1}\right)}^n \rmd z ,
\end{align*}
 the ``perturbed integrating superoperator'', and the ``perturbed eigen-nilpotent'':
\begin{equation*}
\Ss_{\hat H} = {1\over{2\pi \hbar}}\oint_{|z|=\epsilon} z^{-1}\Rs_{\hat H(z)}\rmd z,\qquad \Ds_{\hat H} = {\hbar\over{2\pi}}\oint_{|z|=\epsilon} z\,\Rs_{\hat H(z)}\rmd z .
\end{equation*}
Only  coefficients of $z^{-1}$ in these expansions will contribute to the result:
\begin{align}
\Ps_{\hat H}&=\sum_{n=0}^\infty{(-1)}^{n+1}\alpha^n\left(\sum_{
\sum{p_j}={\bf n}\atop p_j\ge 0}\Srs_{\hat H_0}^{(p_{n+1})}\underbrace{\Ls_{\Hi}\Srs_{\hat H_0}^{(p_n)}\,\ldots\,\Srs_{\hat H_0}^{(p_2)}
\Ls_{\Hi}}_{n\ {\rm times}}\Srs_{\hat H_0}^{(p_1)}\right),\nonumber\\
\Ss_{\hat H}&=\sum_{n=0}^\infty{(-1)}^{n\hphantom{+1}}\alpha^n\left(\sum_{
\sum{p_j}={\bf n+1}\atop p_j\ge 0}\Srs_{\hat H_0}^{(p_{n+1})}\underbrace{\Ls_{\Hi}\Srs_{\hat H_0}^{(p_n)}\,\ldots\,\Srs_{\hat H_0}^{(p_2)}
\Ls_{\Hi}}_{n\ {\rm times}}\Srs_{\hat H_0}^{(p_1)}\right),\label{Pexp}\\
\Ds_{\hat H}&=\sum_{n=1}^\infty{(-1)}^{n+1}\alpha^n\left(\sum_{
\sum{p_j}={\bf n-1}\atop p_j\ge 0}\Srs_{\hat H_0}^{(p_{n+1})}\underbrace{\Ls_{\Hi}\Srs_{\hat H_0}^{(p_n)}\,\ldots\,\Srs_{\hat H_0}^{(p_2)}
\Ls_{\Hi}}_{n\ {\rm times}}\Srs_{\hat H_0}^{(p_1)}\right).\nonumber
\end{align}

Because the superoperator $\Srs_{\hat H_0}^{(p_j)}$ consists of $p_j$ superoperators  $\Ss_{\hat H_0}$,
the summation in the above expressions should be done by all possible placements  of $n$ (or $n+1$, or  $n-1$)  
$\Ss_{\hat H_0}$   in $n+1$ sets. 
There are $C^n_{2n}$ terms of order $\alpha^n$  in $\Ps_{\hat H}$  and $C^n_{2n+1}$ terms in $\Ss_{\hat H}$.
For the Hermitian $\hat G$, the operators $\Ss_{\hat H} \hat G$ and $\Ds_{\hat H} \hat G$ are anti-Hermitian.

The properties of unperturbed superoperators can be  extended to their analytic continuations  
as follows:
\[\Ps_{\hat H} {\hat H} = {\hat H}, \quad\Ss_{\hat H}\Ls_{\hat H} = 1-\Ps_{\hat H},\quad\Ls_{\hat H} \Ps_{\hat H} = \Ps_{\hat H} \Ls_{\hat H} = \Ds_{\hat H}, \, {\textrm{etc.}}\]
For the details, see  our work \cite{nikolaev4}  on an application of the Kato series to 
classical mechanics.

To avoid misunderstanding, it should be noted that $\Ps_{\hat H} \hat F$ will not be  commuting with  the perturbed Hamiltonian. 
This is because $\Ls_{\hat H} \Ps_{\hat H} =  \Ds_{\hat H}\neq 0$ in general. 
Actually, the  ``perturbed superprojector''  $\Ps_{\hat H}$ projects onto some analytic continuation
of the algebra of integrals of the unperturbed Hamiltonian. 
This may not coincide, in general, with the algebra of integrals of the perturbed system, because of a destruction of  symmetries.
 In other words, the zero eigenvalue of the Liouville superoperator may be split by the perturbation.

\subsection{Canonical properties of the  Liouvillian superresolvent}

Because  $\Ls$ is a derivative, there  exists an  { integration by parts} formula for its superresolvent \cite{nikolaev4}.
For any $z_1$, $z_2$, and $z_3$  outside of the spectrum  of $\Ls_{\hat H}$ and any operator $\hat F$,  the following holds true:
\begin{align}
\Rs_{\hat H}(z_1) \Ls_{\Rs_{\hat H}(z_2) \hat F} -& \Ls_{\Rs_{\hat H}(z_2) \hat F} \,\Rs_{\hat H}(z_3) 
 + \Rs_{\hat H}(z_1) \Ls_{\hat F} \, \Rs_{\hat H}(z_3)\nonumber\\
&= (z_1-z_2-z_3)\, \Rs_{\hat H}(z_1) \Ls_{\Rs_{\hat H}(z_2) \hat F}\,  \Rs_{\hat H}(z_3).\label{SimpID1}
\end{align}
This  results from the  application  of the identical superoperator
$
\Rs_{\hat H}(z_1)(\ih\Ls_{\hat H} - z_1) \equiv  1
$ 
 to the commutator $\Ls_{\Rs_{\hat H}(z_2) \hat F}\,\Rs_{\hat H}(z_3)$ 
and the expansion of the Jacobi identity. 

Consider the derivative of the Liouvillian superresolvent  with respect to the perturbation
\[
\frac{\partial}{\partial\alpha}\Rs_{\hat H}(z) = -\ih\,\Rs_{\hat H}(z)\,\Ls_{\Hi}\,\Rs_{\hat H}(z).
\]
Substituting $z_1=z_3=z$, and $\hat F=\Hi$  into the  canonical superresolvent 
identity $(\ref{SimpID1})$ 
 yields
\begin{equation}
\Rs_{\hat H}(z)\,\Ls_{\Hi}\,\Rs_{\hat H}(z) = \Ls_{\Rs_{\hat H}(z_2) \Hi}\Rs_{\hat H}(z) - \Rs_{\hat H}(z)\Ls_{\Rs_{\hat H}(z_2) \Hi} - z_2\,  \Rs_{\hat H}(z) \Ls_{\Rs_{\hat H}(z_2) \Hi}\,\Rs_{\hat H}(z). \label{SimpID2}
\end{equation}\normalsize
Look at the coefficient of $z_2^0$ in the Laurent series of the above  expression:
\[
\frac{\partial}{\partial\alpha}\Rs_{\hat H}(z) = 
\Rs_{\hat H}(z)\Ls_{\Ss_{\hat H} \Hi} - \Ls_{\Ss_{\hat H}\, \Hi}\Rs_{\hat H}(z) - \ih\, \Rs_{\hat H}(z)\Ls_{\Ps_{\hat H} \Hi}\,\Rs_{\hat H}(z).
\]
Proceeding similarly for coefficients of $z_2^{-n}$ \hbox{$(n\ge1)$} in $(\ref{SimpID2})$, 
we obtain 
\begin{align*}
\Rs_{\hat H}(z)\Ls_{\Ps_{\hat H} \Hi}&=\Ls_{\Ps_{\hat H} \Hi}\,\Rs_{\hat H}(z) - \ih\,\Rs_{\hat H}(z)\Ls_{\Ds_{\hat H} 
\Hi}\,\Rs_{\hat H}(z),\\
\Rs_{\hat H}(z)\Ls_{\Ds^n_{\hat H} \Hi}&=\Ls_{\Ds^n_{\hat H} \Hi}\,\Rs_{\hat H}(z) - \ih\,\Rs_{\hat H}(z)\Ls_{\Ds^{n+1}_{\hat H} \Hi}\,\Rs_{\hat H}(z).
\end{align*}
This allows  us to rewrite the expression for  the  superresolvent derivative as
\begin{align*}
{\partial\over \partial\alpha}\Rs_{\hat H}(z) =&
\,\Rs_{\hat H}(z)\Ls_{\Ss_{\hat H} \Hi} - \Ls_{\Ss_{\hat H} \Hi}\Rs_{\hat H}(z) \\
&- \ih\Ls_{\Ps_{\hat H} \Hi}\Rs_{\hat H}(z)^2 
+ \left(\ih\right)^2 \Ls_{\Ds_{\hat H} \Hi}\Rs_{\hat H}(z)^3
- \left(\ih\right)^3\Ls_{\Ds^2_{\hat H} \Hi}\Rs_{\hat H}(z)^4 + \ldots
\end{align*}
Actually, this is a power series because $\Ds^n_{\hat H} = \Ord(\alpha^n)$.

It follows from the Hilbert resolvent identity that
\[
{\partial^n\over \partial z^n}\,\Rs_{\hat H}(z) = n!\ \Rs_{\hat H}^{n+1}(z).
\]
Finally, we obtain
\begin{align}
\frac{\partial}{\partial\alpha}\Rs_{\hat H}(z) =&
\,\Rs_{\hat H}\Ls_{\Ss_{\hat H} \Hi} - \Ls_{\Ss_{\hat H} \Hi}\Rs_{\hat H}(z) 
-\ih \Ls_{\Ps_{\hat H} \Hi}\frac{\partial\,\Rs_{\hat H}(z)}{\partial\,z} \nonumber\\
&{}+ \frac{1}{2}\left(\ih\right)^2 \Ls_{\Ds_{\hat H} \Hi}\frac{\partial^2\Rs_{\hat H}(z)}{\partial\,z^2}
- \frac{1}{6}\left(\ih\right)^3 \Ls_{\Ds_{\hat H}^2 \Hi}\frac{\partial^3\Rs_{\hat H}(z)}{\partial\,z^3} 
+ \ldots
\end{align}
The derivative of superprojector $({\partial}/{\partial\,\alpha})\Ps_{\hat H}$ can be obtained as the residue of this expression at $z=0$.
In our  case of an isolated point spectrum,  the superresolvent  is a meromorphic function, and therefore the residue of any of its derivatives with respect to $z$  vanishes identically. 
As a result, the superprojector $\Ps_{\hat H}$  transforms canonically under  perturbation:
\begin{equation}
\frac{\partial}{\partial\,\alpha}\Ps_{\hat H} = \Ps_{\hat H}\Ls_{\Ss_{\hat H} \Hi} - \Ls_{\Ss_{\hat H} \Hi}\Ps_{\hat H}.\label{Projeq}
\end{equation}
This identity holds to all perturbation orders.

\section{An explicit expression for a  generator}
\subsection{ Ordered exponentials}
Ordered exponentials were introduced in quantum field  theory by Dyson \cite{Dyson49} in 1949. They 
   parametrize an $\alpha$-dependent unitary transformation $\Uo(\alpha)$
\[
\ket{\psi(0)}=\Uo\,\ket{\widetilde\psi(\alpha)},\qquad
\hat{\widetilde{H}} =  \Uo^{-1}\,\hat H\,\Uo,  
\]
and its inverse $\Vo=\Uo^{-1}$ using direct and inverse  Hamiltonian flows in  ``time'' $\alpha$
 with some Hermitian generator $ \hat G(\alpha)$:
\[
\frac{\partial}{\partial\alpha}\Uo = \rmi \,\hat G(\alpha)\, \Uo,\qquad\frac{\partial}{\partial\alpha}\, \Vo = - \rmi\, \Vo \hat G.\]
In  classical mechanics the ordered exponentials are known as  Lie-Deprit transforms \cite{Deprit}.
Their applications to quantum canonical perturbation theory are discussed by Ali \cite{Ali85} and Scherer \cite{Scherer}. 
We   adopt the  notation by Suzuki  \cite{Suzuki85} for the direct and inverse ordered exponentials: 
\[\Uo=\expp{ \hat G},\qquad \Vo=\expm{\hat G}.
\]
The integral signs here   
have only notational meaning as a reminiscence of the Dyson series.

We will mostly use
the superoperatorial ordered exponentials  that satisfy the equations:
\begin{equation}
\frac{\partial}{\partial\alpha}\Us(\alpha) = \rmi \,\Ls_{\hat G(\alpha)}\, \Us,\qquad
\frac{\partial}{\partial\alpha}\, \Vs(\alpha) = - \rmi\, \Vs \Ls_{\hat G}\ .\label{Deprit}
\end{equation}

  Analogous  to the Hausdorff identity, 
they are  factorizable as follows  \cite{Araki1973}:
\begin{align*} 
\Us(\alpha)\hat F=& \expp{ \Ls_{\hat G}}\ \,\hat F=\expp{ \hat G}\ \,\hat F\,\, \expm{ \hat G},\\
\Vs(\alpha)\hat F=&\expm{  \Ls_{\hat G}} \hat F=\expm{ \hat G} \,\hat F\,\, \expp{ \hat G}. 
\end{align*}
Therefore, the transformed Hamiltonian can be written as
$\hat{\widetilde{H}}=    \expm{  \Ls_{\hat G}}\ \hat H$.

In quantum physics the ordered exponentials 
are usually  computed by the  Dyson series \cite{Dyson49}. 
But canonical perturbation theory  uses their representation as a power series:  
\[\expp{ \Ls_{\hat G}}=\sum_{n=0}^\infty \alpha^n \Us_n,\qquad
\expm{ \Ls_{\hat G}}=\sum_{n=0}^\infty \alpha^n \Vs_n.
\] 
Substituting these   series  and $\hat G(\alpha) = \sum_{n=0}^\infty \alpha^n \hat  G_n$  into  the   equations \eref{Deprit}, 
Deprit   
obtained   the   relations for the coefficients: 
\begin{equation}
\Us_n = \frac{\rmi }{n} \sum_{k=0}^{n-1} \Ls_{\hat {G}_{n-k-1}} \Us_k, \qquad
\Vs_n = -  \frac{\rmi }{n} \sum_{k=0}^{n-1} \Vs_k \Ls_{\hat {G}_{n-k-1}}, \label{Deprit1}
\end{equation}
and  developed the  ``triangular'' computational algorithm \cite{Deprit}.

Iterations of \eref{Deprit1} 
yield non-recursive formulas \cite{Cary}:
\begin{align*}
\Us_n&= \sum_{\substack{(m_1,\ldots,m_r)\\n>m_1>m_2>\cdots>m_r}}\, \rmi^{r+1}\frac{\Ls_{\hat G_{n-m_1-1}}}{n} \frac{\Ls_{\hat G_{m_1-m_2-1}}}{m_1}\cdots\frac{\Ls_{\hat G_{m_r-1}}}{m_r}\ ,\\
\Vs_n&= \sum_{\substack{(m_1,\ldots,m_r)\\n>m_1>m_2>\cdots>m_r}} 
\!\!\!\!\!\!(-\rmi)^{r+1}\frac{\Ls_{\hat G_{m_r-1}}}{m_r} \cdots\frac{\Ls_{\hat G_{m_1-m_2-1}}}{m_1}\frac{\Ls_{\hat G_{n-m_1-1}}}{n}\ .
\end{align*}
Here, the sum runs  over all sets of 
integers $(m_1,\ldots,m_r)$, satisfying $n>m_1>\cdots>m_r>0$.

In  the first  orders:
\begin{align*}
\expp{\Ls_{\hat G}}\   =& 1 + \rmi\,\alpha \Ls_{\hat G_0} - \frac{\alpha^2}{2}(\Ls_{\hat G_0}^2 - \rmi \,\Ls_{\hat G_1})\\
 &-  \rmi\, \frac{\alpha^3}{6} ( \Ls_{\hat G_0}^3 -\rmi\,  \Ls_{\hat G_0} \Ls_{\hat G_1} - 2\rmi \, \Ls_{\hat G_1} \Ls_{\hat G_0} - 2   \Ls_{\hat G_2}) 
+  \Ord(\alpha^3),\\
\expm{\Ls_{\hat G}} =& 1 -\rmi\,\alpha \Ls_{\hat G_0} - \frac{\alpha^2}{2}(\Ls_{\hat G_0}^2 + \rmi \Ls_{\hat G_1})\\
&+ \rmi \,\frac{\alpha^3}{6} ( \Ls_{\hat G_0}^3 +
 2\rmi\,\Ls_{\hat G_0} \Ls_{\hat G_1} +\rmi\,  \Ls_{\hat G_1} \Ls_{\hat G_0} 
-  2 \Ls_{\hat G_2}) +  \Ord(\alpha^3).
\end{align*}
\normalsize

\subsection{Intertwining transformation }

The canonical  identity \eref{Projeq} means  that  
the ordered exponential  
with the generator $\rmi\,\Ss_{\hat H}  \Hi$:
\begin{align*}\ket{\psi(0)}&=\expp{\rmi\,\Ss_{\hat H}  \Hi}\ \ket{\widetilde\psi(\alpha)},\\
\hat{\widetilde{H}}\ \ \ \ \ & 
  = \expm{  \Ls_{\rmi\,\Ss_{\hat H}  \Hi}}\ \hat H,
\end{align*}
unitarily connects the perturbed and unperturbed superprojectors:
\[
\Ps_{\hat H} = {\expp{\Ls_{\rmi\,\Ss_{\hat H}  \Hi}}\, \Ps_{\hat H_0}\, \expm{\Ls_{\rmi\,\Ss_{\hat H} \Hi}}}.
\]

Since this transformation intertwines the  superprojectors
\begin{equation}
{\expm{\Ls_{\rmi\,\Ss_{\hat H}  \Hi}}\,\Ps_{\hat H} =  \Ps_{\hat H_0}\, \expm{\Ls_{\rmi\,\Ss_{\hat H} \Hi}}},\label{inter}
\end{equation}
and
  $\Ps_{\hat H} {\hat H} = {\hat H}$, the transformed Hamiltonian  becomes block-diagonal:
\[
\hat{\widetilde{H}} =  \expm{\Ls_{\rmi\,\Ss_{\hat H} \Hi}}\, \Ps_{\hat H}\hat H 
= \Ps_{\hat H_0}\,\expm{\Ls_{\rmi\,\Ss_{\hat H} \Hi}}\, \hat H  
 = \Ps_{\hat H_0}\hat{\widetilde{H}} .
\]

We have thus shown   that  the ordered exponential  
with the generator
\begin{multline}
\hat G=\rmi\,\Ss_{\hat H} \Hi=\rmi\,\sum_{n=0}^\infty{(-1)}^{n+1}\alpha^n\left(\sum_{
\sum
{p_j}={\bf n+1}\atop p_j\ge 0}\Srs_{\hat H_0}^{(p_{n+1})}
\underbrace{\Ls_{\Hi}\Srs_{\hat H_0}^{(p_n)}\,\ldots\,\Srs_{\hat H_0}^{(p_2)}
\Ls_{\Hi}}_{n\ {\rm times}}\Srs_{\hat H_0}^{(p_1)} \Hi
\right) \label{explicit}
\end{multline}
formally block-diagonalizes the Hamiltonian in all orders in $\alpha$. 

The above formula can be extended to 
  systems  with degenerate  energy levels, or to unbounded perturbation in the form of an
asymptotic series. 
It may be shown \cite{nikolaev4} that an expression for $\hat G$ truncated at order $N$   results in a block-diagonal Hamiltonian up to $\Ord(\alpha^{N+1})$.
For  clarity, we   here use formal ``analytic''  expressions. But for degenerate systems,
it must  always be remembered that  these expressions must be converted straightforwardly  into 
 truncated sums,  and all  equalities hold up to   $\Ord(\alpha^{N+1})$.

In the first perturbative orders, the  generator
\begin{multline*}
\hat G=\rmi\,\Ssh \Hi - \rmi\alpha\left( \Ssh\Lsi\Ssh - \Ssh^2\Lsi\Psh - \Psh\Lsi\Ssh^2 \right) \Hi \\
{}     + \rmi\alpha^2  \left(\Ssh\Lsi\Ssh\Lsi\Ssh  
{} - \Ssh\Lsi\Ssh^2\Lsi\Psh - \Ssh\Lsi\Psh\Lsi\Ssh^2 -\Ssh^2\Lsi\Ssh\Lsi\Psh
                  \right.\\
{}\qquad
 - \Ssh^2\Lsi\Psh\Lsi\Ssh  -\Psh\Lsi\Ssh\Lsi\Ssh^2 - \Ps\Lsi\Ssh^2\Lsi\Ssh +\Ssh^3\Lsi\Psh\Lsi\Psh  
                   \\
{}\left. + \Psh\Lsi\Ssh^3\Lsi\Psh + \Psh\Lsi\Psh\Lsi\Ssh^3 \right) \Hi + \Ord(\alpha^3)
\end{multline*}
yields  the following block-diagonal effective Hamiltonian up to $\alpha^4$ :
\begin{multline}
\hat{\widetilde{H}}=\hat H_0 
       + \alpha \Psh \Hi 
       -  \frac{\alpha^2}{2} \Psh\Lsi\Ssh   \Hi 
       + \alpha^3 \left(\frac{1}{3} \Psh\Lsi\Ssh\Lsi\Ssh 
                 - \frac{1}{6} \Psh\Lsi\Ssh^2\Lsi\Psh \right) \Hi \\
      {}+\alpha^4 \left(\frac{1}{6}\Psh\Lsi\Ssh\Lsi\Ssh^2\Lsi\Psh  
                 {}- \frac{1}{4}  \Psh\Lsi\Ssh\Lsi\Ssh\Lsi\Ssh   
                 + \frac{1}{12} \Psh\Lsi\Ssh^2\Lsi\Ssh\Lsi\Psh \right.\\
               {}+ \frac{1}{8} \Psh\Lsi\Ssh^2\Lsi\Psh\Lsi\Ssh   
               {}+ \frac{1}{4} \Psh\Lsi\Psh\Lsi\Ssh^2\Lsi\Ssh 
                 + \frac{1}{4} \Psh\Lsi\Psh\Lsi\Ssh\Lsi\Ssh^2  \\
                 - \frac{1}{6} \Psh\Lsi\Psh\Lsi\Ssh^3\Lsi\Psh  
  {}\left.               - \frac{1}{4} \Psh\Lsi\Psh\Lsi\Psh\Lsi\Ssh^3\right)  \Hi + \Ord(\alpha^5).\label{KatoGen}
\end{multline}
Here  we  omitted the indices $\hat H_0$ for unperturbed operators $\Ps_{\hat H_0}$ and $\Ss_{\hat H_0}$. 
These expressions  are more regular than corresponding formulas \cite{Tyuterev,Klein,Shavitt} for  standard CPT methods. 

\subsection{General form of the generator}
Knowing that $\Ps_{\hat H_0}$ and $\Ps_{\hat H}$ are unitary connected  and that this transformation block-diagonalizes the Hamiltonian, 
the procedure of canonical perturbation theory can be viewed as 
{\em the construction of  unitary transformation, which connects unperturbed and perturbed  superprojectors.}

Let us now determine the general form of the generator of
such  a transformation. The perturbed $\Ps_{\hat H}$ should satisfy  the differential equation
\[
\frac{\partial}{\partial\,\alpha}\Ps_{\hat H} = \rmi \Ls_{\hat G}\, \Ps_{\hat H} - \rmi\, \Ps_{\hat H}\, \Ls_{\hat G}.
\]
Application of  this expression  to the Hamiltonian $\hat H$  and the identities  
\[\Ps_{\hat H} {\hat H} = {\hat H},\qquad  
{\partial\over \partial\,\alpha}(\Ps_{\hat H} {\hat H}) =
 \left({\partial\over \partial\,\alpha}\Ps_{\hat H} \right)  {\hat H} + 
\Ps_{\hat H}  \frac{\partial}{\partial\,\alpha}  {\hat H},\qquad {\partial\over \partial\,\alpha} {\hat H} = \Hi,
\]
yield
\[
(1-\Ps_{\hat H}) \Ls_{\hat H} \hat  G = \rmi\, (1-\Ps_{\hat H}) \Hi.
\]
To solve this equation, it is sufficient to apply the $\Ss_{\hat H}$ superoperator. Therefore, the general form of the generator of intertwining unitary  transformation is
\begin{equation}
\hat G = \rmi\, \Ss_{\hat H} \Hi + \Ps_{\hat H} \hat F,\label{Final}
\end{equation}
where $\hat F(\alpha)$  may be any  Hermitian operator. This is the main result of this work. This formula provides a non-recursive expression for
the generator of the block-diagonalizing transformation and defines its ambiguity. 

The  choice of  $\Ps_{\hat H} \hat F$ is the {\em uniqueness  condition} \cite{soliverez80}. It is natural to choose $\hat F\equiv 0$ or $\Ps_{\hat H} \hat G = 0$.
This is not equal to the ``off-diagonal'' condition $\Ps_{\hat H_0} \hat G_P = 0$  traditionally used  in  canonical perturbation theory \cite{Primas}. 
Because $\Ls_{\Ps_{\hat H} \hat F}\, \Ps_{\hat H} = \Ps_{\hat H} \Ls_{\Ps_{\hat H} \hat F}$, the superprojector $\Ps_{\hat H}$  is itself insensitive to this choice.

We can conclude that {\em the generators of block-diagonalizing transformations 
can differ by a function belonging to  a continuation of  the algebra of integrals of unperturbed system.}

{Effective Hamiltonians} that were  block-diagonalized using different uniqueness conditions $ \hat F_1$ and $ \hat F_2$
 are connected by unitary transformation:
\[ \Us_{21} = \expm{\Ls_{\hat G_2}} \ \ \expp{\Ls_{\hat G_1}}.
\]
Let us find its generator:  
\begin{align*} \frac{\partial}{\partial\alpha}\Us_{21} &= \rmi\ \expm{\Ls_{\hat G_2}}
\ \left(\Ls_{\hat G_1}-\Ls_{\hat  G_2}\right)\ \expp{\Ls_{\hat G_1}}
\\
&= \rmi\,\expm{\Ls_{\hat G_2}}\Ls_{\Ps_{\hat H} (\hat F_1 - \hat F_2)}\expp{\Ls_{\hat G_1}}\\
&= \rmi\,\Ls_{\expm{\Ls_{\hat G_2}}\Ps_{\hat H} (\hat F_1 - \hat F_2)}\Us_{21}.
\end{align*}
Because of the intertwining  relation \eref{inter},  the generator of  $\Us_{21}$ is  always block-diagonal:
\[\hat G_{21}=\expm{\Ls_{\hat G_2}}\, \Ps_{\hat H} (\hat F_1 - \hat F_2) =
 \Ps_{\hat H_0} \expm{\Ls_{\rmi\, \Ss_{\hat H} \Hi + \Ps_{\hat H} \hat F_2}} \, (\hat F_1 - \hat F_2).
\]
This is  the well-known result 
that
{  the effective Hamiltonian is determined up to block-diagonal transformation} \cite{soliverez81}. 
Such ambiguity does not affect the perturbed 
eigenvalues.

\section{Computational aspects}
A major difference between this and  standard canonical perturbation algorithms by Van Vleck \cite{VanVleck}, Primas \cite{Primas},  etc.,
is the  explicit non-recursive formulas.  Traditionally, 
perturbation  computations block-diagonalize the Hamiltonian order by order.
In contrast, we    compute  the  generator $\hat G = \rmi\, \Ss_{\hat H}  \Hi$ up to  the  
desired order directly. 
Then, the  ordered exponential block-diagonalizes the Hamiltonian.
\subsection{An explicit algorithm for the  generator}  
Let us introduce the  superoperators
\[
\Zs_n^m = {(-1)}^{n+1}\sum\limits_{
p_1+\,\ldots\,+p_{n+1}=m\atop p_j\ge 0}\Srs_{H_0}^{(p_{n+1})}
\underbrace{\Ls_{\Hi}\Srs_{H_0}^{(p_n)}\,\ldots\,\Srs_{H_0}^{(p_2)}
\Ls_{\Hi}}_{n\, {\rm times}}\Srs_{H_0}^{(p_1)}.
\]
Here, the sum runs over all possible placements  of $m$  superoperators $\Ss_{H_0}$  in $n+1$ sets. 
The simplest such expressions are $\Zs_0^0=\Ps_{H_0}$, and  $\Zs_0^m=-\Ss_{H_0}^m$ for $m>0$. 
We have  already discussed their computation in section \ref{sec:basic}.

The following recursive relation  
\[
\Zs_n^m = \sum_{k=0}^m \Zs_{0}^{m-k} \Ls_{\Hi} \Zs_{n-1}^k
\]
 leads to an efficient  algorithm for the generator $\hat G=\rmi\, \Ss_H  \Hi$.  
We consider the square table:
\[
\begin{array}{llllll}
 \hat F^{0}_{0}       & \hat {\bm F^{1}_{0}} &\hat F^{2}_{0}&\ldots&\hat F^{N}_{0}&\hat F^{N+1}_{0}\\
 \hat F^{0}_{1}       &\hat F^{1}_{1}  &\hat {\bm F^{2}_{1}}&\ldots&\hat F^{N}_{1}&\hat F^{N+1}_{1}\\
&\ldots&&&\\
\hat F^{0}_{N-1} &\hat F^{1}_{N-1}    &\hat F^{2}_{N-1}&\ldots&\hat {\bm F^{N}_{N-1}}&\hat F^{N+1}_{N-1}\\
&&&&&\hat  {\bm F^{N+1}_{N}}
\end{array}
\]
Here the first row is 
\[\hat F_0^0 = \Zs_0^0 \Hi=\Ps_{H_0} \Hi, \qquad \hat F_0^m({\bf x}) =\Zs_0^m \Hi=-\Ss_{H_0}^m \Hi,\]
and  each subsequent row is generated from the previous row according to the rule
\[\hat F_{n+1}^m  = \sum_{k=0}^m \Zs_{0}^{m-k} \Ls_{\Hi} \hat F_{n}^k. \]
 The generator truncated at order  $\alpha^{N+1}$  is given by $\hat G_{[N]}= \rmi\sum_{n=0}^N \alpha^n \hat F_n^{n+1}$. 

\subsection{Computation of the ordered exponential}  
In classical mechanics the ordered exponentials are usually computed using the Deprit ``triangle'' \cite{Deprit}.
However,  
we have  observed that for all of our quantum examples 
the following algorithm is faster.

Up to $\Ord(\alpha^{N+1})$ the transform of general perturbed Hamiltonian  $\hat H=\sum_{n=0}^N \alpha^n \hat H_n$ can be written
 in the following form:
\[
\hat{\widetilde H}=  \expm{\Ls_{\hat G}} \hat H = \left( \sum_{n=0}^N \alpha^n\, \Vs_n \right) \hat H 
= \sum_{n=0}^N  \Vs_n \hat{\widetilde F}{}_{n}^{(N)}.
\]
Here we introduced ancillary double-indexed operators $\hat{\widetilde F}{}^{(N)}_{n} = \sum_{k=n}^{N} \alpha^{k} \hat H_{k-n}$.

The  Deprit relation \eref{Deprit1} 
allows us to express repeatedly  the superoperators $\Vs_n$ by means of 
$\Vs_{\textrm{less then }n}$: 
\[
\hat{\widetilde H}= 
 \Vs_N\hat{\widetilde F}{}_{N}^{(N)}+ \sum_{k=0}^{N-1}  \Vs_k \hat{\widetilde F}{}_{k}^{(N)} = 
 \sum_{k=0}^{N-1}  \Vs_k \hat{\widetilde F}{}_{k}^{(N-1)}
 =\ldots= \sum_{k=0}^{n}  \Vs_k \hat{\widetilde F}{}_{k}^{(n)}=\ldots ,
\]
where  the operators $\hat{\widetilde F}{}_{k}^{(n)}$, $n=N-1,\ldots,0$ are computed using the following relation:
\[
\hat{\widetilde F}{}_{k}^{(n)} =
 \hat{\widetilde F}{}_{k}^{(n+1)} - \frac{\rmi}{n+1} \Ls_{\hat G_{n-k}} \hat{\widetilde F}{}_{n+1}^{(n+1)},
 \qquad  k=0,\ldots,n.
\]
 Finally, all the superoperators $\Vs$ disappear and we obtain  the transformed Hamiltonian:
\[
\hat{\widetilde H} =
\hat{\widetilde F}{}_{0}^{(0)}.
\]

\subsection{Examples and comparison with other methods}
The explicit  expression is important from a general mathematical standpoint because   
it systematizes and simplifies the perturbation expansion. 
Moreover, 
it provides a sufficiently effective computational
algorithm.
Here we compare   the computer times of  block-diagonalization 
 using the explicit expression \eref{explicit} 
 with those of the Van Vleck  \cite{VanVleck} and Magnus \cite{Magnus,Blanes2009151} approaches.

The Van Vleck  method block-diagonalizes the Hamiltonian using a chain of exponents: 
\[
\hat{\widetilde{H}} = \rme^{-\rmi \alpha^n \hat G_{n-1}}\ldots \rme^{-\rmi \alpha\, \hat G_0}\, \hat H\,  \rme^{\rmi\alpha\, \hat G_0}
\ldots \rme^{\rmi \alpha^n \hat G_{n-1}}\\=
\rme^{- \rmi \alpha^n \Ls_{\hat  G_{n-1}}} \ldots \rme^{- \rmi \alpha \,\Ls_{\hat  G_0}} \hat  H. 
\]
In  classical mechanics it  is  known as  the Dragt-Finn transformation \cite{DragtFinn}.

The Magnus method uses the superoperator exponent  with an $\alpha$-dependent generator \cite{Primas}:
\[
\hat{\widetilde{H}}\vphantom{H} = \rme^{-\rmi \sum_1^n \alpha^k \hat G_{k-1}}\, \hat H  \,
 \rme^{\rmi   \sum_1^n \alpha^k \hat G_{k-1}} =
 \rme^{- \rmi   \sum_1^n \alpha^k \Ls_{\hat  G_{k-1}}} \hat  H .
\]
In  classical mechanics this    is  known as  the Hori algorithm \cite{Hori}.

We have  evaluated the efficiency of these methods for
 the following model systems:
\begin{itemize}
\item The one-dimensional oscillator with the quartic anharmonicity:
\[\hat H=\frac{1}{2}(\hat p^2+\hat q^2)+\frac{\alpha}{4} \hat q^4.
\]
 The  first orders of 
the block-diagonalized Hamiltonian are:
\begin{align*}
\hat{\widetilde{H}}=& \hbar (\hat N+\frac{1}{2}) 
+ \alpha \hbar^2 \left(  \frac{3}{8}{(\hat N+\frac{1}{2})}^2 + \frac{3}{32} \right) 
- \alpha^2 \hbar^3 \left( \frac{17}{64} {(\hat N+\frac{1}{2})}^3 + \frac{67}{256} (\hat N+\frac{1}{2})   \right)+\Ord(\alpha^3). 
\end{align*}
Here $\hat N=\hat a^\dagger\hat a$. 
This is the typical structure of the 
 perturbation series for nondegenerate systems. It is worth noting that the effective diagonal Hamiltonians of such   systems are identical for all the CPT methods.

\item{The H\'enon-Heiles system.}  This is a two-dimensional system with the Hamiltonian
\[\hat H= \frac{1}{2} (\hat  p_1^2+ \hat q_1^2 + \hat p_2^2+\hat q_2^2)+ \alpha (\hat q_1^2 \hat q_2-\frac{1}{3} \hat q_2^3).
\]

The block-diagonalized effective Hamiltonian is:
\begin{align*}
\hat{\widetilde{H}}=& \left(\hbar -  \frac{1}{9} \alpha^2\hbar^2 - \frac{11}{108}\alpha^4 \hbar^3
 \right)  + \left(\hbar -  \frac{2}{3}\alpha^2\hbar^2 - \frac{61}{54}\alpha^4\hbar^3\right)
         \left(\xac{}{1}\xa{}{1} + \xac{}{2}\xa{}{2}\right) \\&{}
- \left(  \frac{5}{12} \alpha^2\hbar^2 + \frac{47}{48}\alpha^4 \hbar^3 \right) 
            \left(\xac{2}{1}\xa{2}{1} + \xac{2}{2}\xa{2}{2}\right)
- \left(\frac{7}{12}  \alpha^2\hbar^2 - \frac{7}{48}\alpha^4 \hbar^3\right) 
            \left(\xac{2}{2}\xa{2}{1} + \xac{2}{1}\xa{2}{2} \right)\\&{}         
+ \left(\frac{1}{3}\alpha^2\hbar^2 -  \frac{9}{4}\alpha^4\hbar^3\right)\xac{}{1}\xa{}{1}\xac{}{2}\xa{}{2}\\&{}
   + \alpha^4\hbar^3\left( 
          \frac{101}{432} \xac{3}{1}\xa{3}{1} 
        - \frac{161}{144} \xac{3}{1}\xa{}{1}\xa{2}{2}
        - \frac{65}{16} \xac{2}{1}\xac{}{2} \xa{2}{1}\xa{}{2} 
       + \frac{175}{144}\xac{2}{1}\xac{}{2}\xa{3}{2} \right.\\&{}\left.\quad
        - \frac{235}{432} \xac{3}{2}\xa{3}{2} 
        - \frac{161}{144}  \xac{}{1}\xac{2}{2}\xa{3}{1}
       + \frac{47}{16}\xac{}{1}\xac{2}{2}\xa{}{1}\xa{2}{2} 
       +  \frac{175}{144}\xac{3}{2} \xa{2}{1}\xa{}{2} 
\right) 
 +\Ord(\alpha^6).
\end{align*}
This expression could be  further diagonalized using finite canonical transformations \cite{nikolaev}.

All three CPT  methods build near-identity unitary block-diagonalizing  transformations.   Such  transformations   
are equivalent to    
 ordered exponentials with different conditions $\Ps_{\hat H} \hat F$.
As a result, all the three effective block-diagonal Hamiltonians of the H\'enon-Heiles system differ from each other starting from the  $8\textsuperscript{th}$ 
perturbation order, and are connected by block-diagonal unitary transformations. 
\end{itemize}

The comparison of efficiency of these methods is not unambiguous. 
High order computations process large multi-gigabyte expressions containing millions of terms.
The performance of operations with such expressions depends
on  the computer algebra system used, its internal  optimizations, the server CPU, RAM, 
 disk subsystem, etc.
Even the relative efficiencies of methods may vary. 
Therefore, the following our results are only illustrative.

\begin{figure}[h!]
\centering
\includegraphics[scale=0.7]{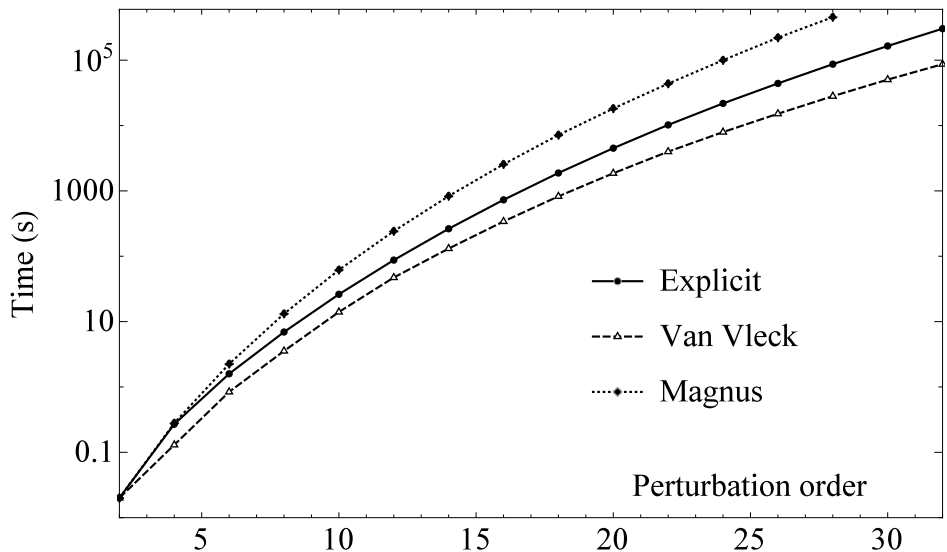}
\vspace{-30pt}
\caption{\label{fig:comp}Block-diagonalization time for H\'enon-Heiles Hamiltonian.}
\end{figure}

Figure \ref{fig:comp} compares 
the computational times of  block-diagonalization for the H\'enon-Heiles system 
on an Oracle\texttrademark{} Exadata X2-2 server with Intel Xeon X5675 (3.06 GHz) processor using Form~4.1 computer algebra system  \cite{Form}.

The computation times for the  quartic anharmonic oscillator  follow the same pattern.
In these particular cases the explicit algorithm is faster  for  high-orders computations than 
the Magnus expansion, but less efficient then the Van Vleck method.

\section{Summary}
We have presented here an application of Kato resolvent expansion to quantum superoperatorial canonical 
perturbation theory. The canonical  identity for the superresolvent  allowed us to  demonstrate   
unitary intertwining of perturbed and unperturbed averaging superprojectors. This leads to 
the  explicit expression  for the  generator of the block-diagonalizing transformation in any perturbation order 
and systematic description of  
  ambiguities in the generator and block-diagonalized  Hamiltonian.
  
  We have also compared the  computational efficiency of the explicit expression 
  for the quartic anharmonic oscillator and the H\'enon-Heiles system
with the  efficiencies of the Van Vleck and Magnus methods up to  the $32\textsuperscript{nd}$ perturbation order.

\begin{acknowledgments}
The author gratefully acknowledges  Professor S.~V.\ Klimenko,
 RDTEX Technical Support Centre Director
  S.~P.\ Misiura, and V.~V.\ Romanova for encouragement and support.
\end{acknowledgments}

\bibliography{kato_vanvleck}

\end{document}